\pdfoutput=1
\documentclass[journal]{IEEEtran}

\usepackage{amsmath,amssymb,amsfonts}
\usepackage{mathtools}
\usepackage{bm}
\usepackage{textcomp}
\usepackage{xcolor}
\usepackage{nicefrac}

\usepackage{graphicx}
\graphicspath{{figures/}{Tikz_figures/}}

\usepackage{subcaption}
\usepackage{booktabs}
\usepackage{multirow}
\usepackage{array}      
\usepackage{tabularx}
\usepackage{colortbl}   

\newcolumntype{L}[1]{>{\raggedright\arraybackslash}p{#1}}
\newcolumntype{C}[1]{>{\centering\arraybackslash}p{#1}}
\newcolumntype{R}[1]{>{\raggedleft\arraybackslash}p{#1}}

\newcolumntype{Z}{>{\hsize=0.95\hsize\centering\arraybackslash}X}          
\newcolumntype{B}{>{\hsize=1.35\hsize\centering\arraybackslash\bfseries}X} 

\usepackage{siunitx}
\usepackage{tikz}
\usepackage{tikz-3dplot}
\usepackage{pgfplots}
\pgfplotsset{compat=1.18}

\usepgfplotslibrary{polar}
\usepgfplotslibrary{groupplots}
\usepgfplotslibrary{fillbetween}

\usetikzlibrary{spy}
\usetikzlibrary{matrix}
\usetikzlibrary{arrows}
\usetikzlibrary{arrows.meta}
\usetikzlibrary{calc}
\usetikzlibrary{positioning}
\usetikzlibrary{angles}
\usetikzlibrary{quotes}
\usetikzlibrary{intersections}
\usetikzlibrary{decorations.pathreplacing}
\usetikzlibrary{shapes.geometric}
\usetikzlibrary{shapes.misc}
\usetikzlibrary{patterns}

\newlength\figureheight
\newlength\figurewidth

\pgfplotsset{
  every tick label/.append style={font=\scriptsize},
  every axis label/.append style={font=\footnotesize},
  every axis title/.append style={font=\footnotesize},
  every axis legend/.append style={font=\scriptsize}
}

\usepackage{circuitikz}

\providecommand{\tikzsetnextfilename}[1]{}

\usepackage{stfloats}

\newcommand{\figref}[1]{Fig.~\ref{#1}}

\newcommand{\tabref}[1]{Table~\ref{#1}}
\newcommand{\secref}[1]{Section~\ref{#1}}

\usepackage{cite}

\usepackage[hidelinks]{hyperref}

\begin{document}

\title{A 28-GHz Varactor-Based RIS With Continuous Phase Control: From Unit-Cell Modeling to Programmable Wavefront Control and Synthesis}

\author{Spandan~Manna, Florian~Reher, Karim~El~Isa,%
        ~\IEEEmembership{Graduate Student Member, IEEE}, 
        \\Amar~Al-Bassam,~\IEEEmembership{Member, IEEE}, 
        and Dirk~Heberling,~\IEEEmembership{Senior Member, IEEE}%
\thanks{This work has been submitted to the IEEE for possible publication. Copyright may be transferred without notice, after which this version may no longer be accessible.}%
\thanks{This work was funded by the German Federal Ministry of Education and Research (BMBF) in the course of the 6GEM research project under grant number 16KISK038. The work of A. Al-Bassam was supported in part by the Fraunhofer Internal Programs
under Grant No. Attract 40-11924. (\textit{Corresponding author: Spandan Manna.})}%
\thanks{S. Manna, F. Reher, and K. El Isa are with the Institute of High Frequency Technology, RWTH Aachen University, 52074 Aachen, Germany (e-mail: manna@ihf.rwth-aachen.de).}%
\thanks{A. Al-Bassam is with the Fraunhofer Institute for High Frequency Physics and Radar Techniques (FHR), 53343 Wachtberg, Germany.}%
\thanks{D. Heberling is with the Institute of High Frequency Technology, RWTH Aachen University, 52074 Aachen, Germany, and also with the Fraunhofer Institute for High Frequency Physics and Radar Techniques (FHR), 53343 Wachtberg, Germany.}%
}

\maketitle

\begin{abstract}
This paper presents a $\SI{28}{GHz}$ varactor-based reconfigurable intelligent surface (RIS) platform with continuous phase control and establishes a unified device-to-system validation framework for programmable electromagnetic wavefront control and synthesis. The proposed RIS comprises \num{96} independently controlled elements, each employing a single varactor diode, with a fully board-integrated analog-bias control architecture. It is built upon an experimentally validated unit-cell model providing approximately $\SI{300}{\degree}$ of continuous reflection-phase tuning under normal incidence. A unified analytical framework incorporating measured horn illumination, finite phase availability, and unit-cell reflection losses is developed to consistently relate device-level characteristics to beamforming performance. The proposed RIS is evaluated through near-field-to-near-field characterization, near-field-to-far-field beam-steering measurements, and far-field-to-far-field wireless-link experiments. The near-field-to-far-field results exhibit close agreement among analytical predictions, full-wave simulations, and measurements, while the far-field-to-far-field response agrees with simulation and a first-order link-budget estimate. Accurate beam steering is demonstrated for all investigated steering angles within $\pm\SI{45}{\degree}$ across three azimuthal planes, with a maximum deviation of approximately $\SI{2}{\degree}$. The complete prototype, including its integrated driver and bias network, draws only $\SI{0.85}{W}$ and achieves an estimated full-aperture reconfiguration time of approximately $\SI{50}{ms}$. Beyond conventional beam steering, the same hardware platform further enables the experimental investigation of 3-bit, 2-bit, and 1-bit phase quantization and programmable multi-beam wavefront synthesis using a common RF aperture and control architecture. Collectively, these results bridge realistic varactor behavior, analytical modeling, and programmable wavefront synthesis, providing a rigorous basis for the development and experimental validation of continuously tunable millimeter-wave RISs.
\end{abstract}


\section{Introduction}\label{sec:Intro}
\IEEEPARstart{R}{econfigurable} intelligent surfaces (RISs) have emerged as a promising technology for realizing programmable electromagnetic environments in future wireless communication systems by enabling dynamic control of electromagnetic wave propagation through software-defined wavefront manipulation~\cite{Basar_IEEEAccess_2019,DiRenzo_IEEEJSAC_2020,ElMossallamy_TCCN_2020,Di_Renzo_PIEEE_2022}. Their ability to adaptively redirect, focus, or shape incident electromagnetic waves provides an attractive means of improving coverage, mitigating blockage, and enhancing spectral efficiency. This capability becomes particularly important at millimeter-wave (mmWave) frequencies, where severe free-space path loss, increased susceptibility to blockage, and highly directional propagation place increasing demands on adaptive control of the electromagnetic environment~\cite{Rappaport_TAP_2013,Rappaport_TAP_2017}. Passive metasurfaces and reflectarrays can improve coverage through tailored wavefront transformations~\cite{Wong_PRX_2018,Sravan_TAP_Part1_2025,Rocca_TAP_2022,Elsherbeni_reflectarray_2018,Manna_Static_MAPCON_2025}; however, their prescribed responses are fixed after fabrication and cannot adapt to time-varying propagation conditions. RISs address this limitation through electronically reconfigurable apertures, whose performance ultimately depends on the tuning mechanism used to realize accurate, dynamically reconfigurable phase distributions~\cite{Hum_TAP_2014}.

Among electronically reconfigurable implementations, PIN-diode-based RISs have become the most mature solution at mmWave frequencies, with numerous experimentally validated prototypes~\cite{Gros_IEEE_openjournal_2021,Liu_TAP_2023,Shamim_IEEE_TAP_2024, Hu_MTT_2025, Shekhawat_OJAP_2025, Caballero_TAP_2025, Peng_TAP_2026}. Their binary switching operation provides a simple and robust hardware architecture. However, extending these implementations toward higher phase resolution generally requires multiple switching devices and increasingly complex biasing networks~\cite{Zhao_OJCS_2024, Wang_TAP_2024, Shamim_TAP_2025}, making the scalable realization of continuously tunable phase control progressively more challenging. Liquid-crystal-based RISs, in contrast, achieve continuous phase control at mmWave frequencies through electrically tunable dielectric materials~\cite{Kim_TAP_2023, Guirado_TAP_2024, Li_TAP_2024, Quintana_TAP_2024, Moon_OJAP_2025}. However, their relatively slow tuning speed and increased insertion loss continue to limit their suitability for highly dynamic beamforming applications. 

These developments illustrate the growing interest in continuous-phase wavefront manipulation while highlighting the trade-offs associated with different tuning mechanisms. Among electronically tunable approaches, varactor diodes provide continuous analog phase control through reverse-bias voltage tuning while requiring negligible dc power, making them attractive for scalable programmable RIS implementations. Nevertheless, experimentally validated varactor-based RIS implementations remain comparatively limited at mmWave frequencies. This is largely due to the increasing influence of varactor parasitics, the need for large-scale independently calibrated bias-control architectures, and the difficulty of maintaining consistency among analytical models, full-wave simulations, and experimental measurements. 

Existing varactor-based RIS research extends from sub-$\SI{6}{GHz}$ to mmWave frequencies. At sub-$\SI{6}{GHz}$, experimentally validated full-array implementations have already been demonstrated~\cite{Liang_TAP_2022, Yang_TAP_2024}. At mmWave frequencies, however, existing studies have primarily focused on full-wave simulations~\cite{da_Silva_FNC_2023,Marianna_EuCAP_2024,MANNA_RIS_MAPCON_2025}, unit-cell characterization~\cite{Ivashina_TAP_2025, Manna_APS_2025}, or limited experimental demonstrations such as phase-only near-field beam steering~\cite{Wolff_JAP_2023}, reflectarray operation without corresponding steered-beam simulation benchmarks~\cite{Fischer_EuCAP_2025}, and channel-binned two-dimensional beam steering~\cite{Rotshild_MDPI_Elec_2024}. Although these studies demonstrate important individual aspects of continuously tunable mmWave RISs, a comprehensive experimental validation framework that consistently connects realistic device behavior, analytical modeling, and measured system-level beamforming performance remains scarce. Without such a framework, the influence of device-level nonidealities on beamforming performance cannot be rigorously quantified or systematically predicted.

To address these challenges, this work presents a $\SI{28}{GHz}$ continuously tunable RIS platform together with a unified device-to-system validation framework. The main contributions of this work are summarized as follows:
\begin{enumerate}
    \item An experimentally validated $\SI{28}{GHz}$ varactor-based unit-cell model, characterized using a modified-waveguide setup, providing approximately $\SI{300}{\degree}$ of continuous reflection-phase tuning with quantified parasitic loss;
    \item A fully integrated $96$-element RIS with on-board analog bias-control circuitry and a modular, daisy-chainable driver architecture that scales beyond the presented prototype;
    \item A unified aperture-field analytical framework (Models~I--IV) that isolates the individual contributions of finite phase availability and unit-cell reflection loss to array-level beamforming performance;
    \item A complete device-to-system validation chain comprising near-field-to-near-field, near-field-to-far-field, and far-field-to-far-field wireless-link measurements;
    \item Demonstration that a single continuously tunable aperture can additionally emulate 3-/2-/1-bit phase quantization and realize controlled beam splitting using a common RF aperture and control architecture.
\end{enumerate}

The remainder of this paper is organized as follows. \secref{sec:UC} presents the design of the varactor-based RIS unit cell, its simulated reflection response under normal and oblique incidence, and its experimental validation using a modified-waveguide measurement setup. \secref{sec:RIS_Model} describes the integrated $96$-element RIS prototype and its analog bias-control architecture, the feed-distance and aperture-efficiency assessment, and the unified aperture-field analytical framework (Models~I--IV) used to relate unit-cell nonidealities to array-level beamforming. \secref{sec:RIS_Meas} reports the experimental validation, comprising near-field-to-near-field characterization, near-field-to-far-field beam-steering measurements in three azimuthal planes, finite-bit (3-/2-/1-bit) phase-quantization emulation, far-field-to-far-field wireless-link assessment, and controlled beam splitting, followed by a comparison with the state of the art. Finally, \secref{sec:Conclusion} concludes the paper.


\section{Design and Characterization of Varactor-Based RIS Unit Cell}\label{sec:UC}

\subsection{Design Concept of RIS Unit Cell}\label{sec:UC_Design}

\begin{figure}[t]
    \centering
        \setlength\figurewidth{.48\textwidth}
        %
        \includegraphics[width=0.900\figurewidth]{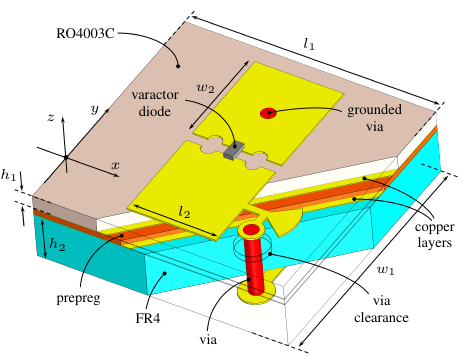}%
        \\[-4pt]
        {\scriptsize (a)}%
        \\[-3pt]
        \begin{minipage}[b]{0.554\figurewidth}%
          \centering
          \parbox[b][0.243\figurewidth][b]{\linewidth}{%
            \centering
            \includegraphics[width=0.842\linewidth]{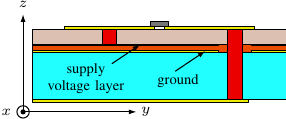}%
          }\\[-4pt]
          {\scriptsize (b)}%
        \end{minipage}%
        \hspace{0.025\figurewidth}%
        \begin{minipage}[b]{0.359\figurewidth}%
          \centering
          \parbox[b][0.243\figurewidth][b]{\linewidth}{%
            \centering
            \includegraphics[width=\linewidth]{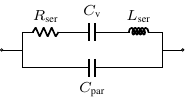}%
          }\\[-4pt]
          {\scriptsize (c)}%
        \end{minipage}%
        \caption{Proposed varactor-loaded RIS unit cell: (a)~geometry and layer configuration, (b)~cross-sectional view of the biasing arrangement, and (c)~equivalent-circuit model of the varactor diode. The unit-cell dimensions are $l_1=w_1=\lambda_0/2=\SI{5.36}{mm}$, $l_2=\SI{1.90}{mm}$, $w_2=\SI{2.19}{mm}$, $h_1=\SI{0.305}{mm}$, and $h_2=\SI{1}{mm}$.}
      \label{Fig:Geometry_Unit_Cell}
    \end{figure}

The proposed RIS unit cell is designed to realize continuous reflection-phase tuning at $\SI{28}{GHz}$ while maintaining a compact footprint compatible with half-wavelength element spacing. The geometry and layer configuration are illustrated in \figref{Fig:Geometry_Unit_Cell} together with the equivalent circuit model of the varactor diode. The unit-cell dimensions are selected as $l_1=w_1=\lambda_0/2=\SI{5.36}{mm}$, enabling array implementation while avoiding grating lobes.

The unit cell is implemented on a multilayer substrate stack comprising a $\SI{0.305}{mm}$ RO4003C substrate ($\varepsilon_r=3.55$, $\tan\delta=0.0027$ at $\SI{10}{GHz}$), an approximately $\SI{0.1}{mm}$ prepreg bonding layer (TU-768P 1080), and a $\SI{1}{mm}$ FR4 biasing substrate ($\varepsilon_r=4.3$, $\tan\delta=0.025$). The radiating structure consists of two rectangular metallic patches separated by a $\SI{0.22}{mm}$ gap and electrically interconnected through a varactor diode. A radial stub is incorporated on the top metallization layer to provide RF isolation between the radiating structure and the bias network while enabling dc biasing of the varactor. The biasing is realized using two vias with a drill diameter of $\SI{0.3}{mm}$ and a clearance diameter of $\SI{0.7}{mm}$. One via connects the bias distribution network to one of the radiating patches, whereas the other connects to the supply voltage layer on the backside of the RO4003C substrate. This arrangement enables independent biasing of each unit cell while preserving the RF performance of the radiating structure.

Particular attention is given to realistic varactor modeling, as parasitic effects become increasingly significant at mmWave frequencies. The employed GaAs flip-chip varactor diode (MAVR-011020-1411, MACOM) is represented using a unit-cell-dependent parasitic model characterized by $L_\mathrm{ser}=\SI{88.5}{pH}$, $C_\mathrm{par}=\SI{9}{fF}$, and $R_\mathrm{ser}=\SI{5.5}{\Omega}$, as reported in~\cite{Manna_APS_2025}. The validity of this model for the modified unit-cell implementation is experimentally reassessed in \secref{sec:UC_Meas}.

\subsection{Unit-Cell Performance Under Normal and Oblique Incidences}\label{sec:UC_Performance}

\begin{figure}[t]
\centering
\subcaptionbox{\hspace*{-28.1pt}\label{Fig:S11_Mag_vs_Freq_Cap_vary}}[0.5\columnwidth][c]{%
    \includegraphics{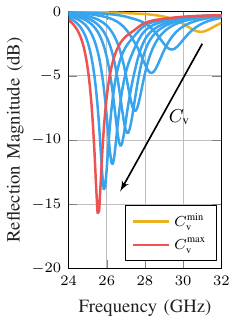}}%
\subcaptionbox{\hspace*{-32.4pt}\label{Fig:S11_Phase_vs_Freq_Cap_vary}}[0.5\columnwidth][c]{%
    \includegraphics{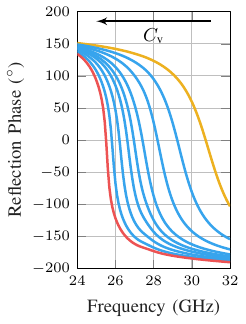}}%
\caption{Simulated reflection response of the proposed RIS unit cell under varactor-based tuning: (a)~magnitude and (b)~phase as the varactor capacitance is varied from $C_\textrm{v}^\textrm{min}=\SI{0.025}{pF}$ to $C_\textrm{v}^\textrm{max}=\SI{0.225}{pF}$.}
\label{Fig:S11_vs_Freq_Cap_vary}
\end{figure}

The reflection characteristics of the proposed unit cell are investigated in CST Studio Suite using a Floquet-port-based full-wave simulation model under periodic boundary conditions, thereby emulating an infinite periodic environment that inherently accounts for mutual coupling between adjacent elements. A linearly polarized electromagnetic wave (along the $y$-direction) is used for excitation. The capacitance of the employed varactor diode is swept over its practical tuning range of $\SIrange{0.025}{0.225}{pF}$.

The simulated reflection magnitude and phase responses are presented in \figref{Fig:S11_vs_Freq_Cap_vary}. Varying the reverse-bias voltage changes the varactor capacitance and consequently the effective surface reactance of the unit cell. As the capacitance increases, the resonance shifts toward lower frequencies, resulting in a continuous variation of the reflection phase. At the target operating frequency of $\SI{28}{GHz}$, the unit cell achieves a phase coverage of approximately $\SI{300}{\degree}$ with a peak reflection loss of $\SI{5.31}{dB}$. 

\begin{figure}[t]
    \centering
    \setlength\figureheight{.45\columnwidth} 
    \setlength\figurewidth{0.70\columnwidth}
    \includegraphics{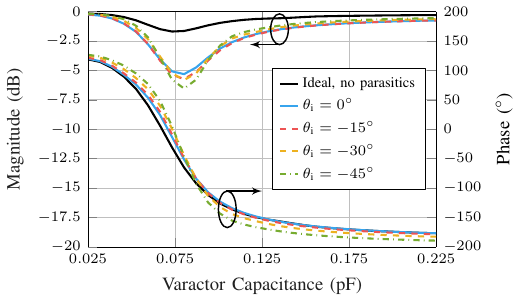}
    \caption{Simulated reflection magnitude and phase versus the varactor capacitance of the proposed RIS unit cell at $\SI{28}{GHz}$ under normal and oblique incidences.}
    \label{fig:UC_Norm_vs_oblique}
    \end{figure}

To isolate the impact of varactor nonidealities, the reflection response is further analyzed as a function of capacitance in \figref{fig:UC_Norm_vs_oblique}. An additional reference simulation was performed by replacing the varactor model [\figref{Fig:Geometry_Unit_Cell}(c)] with an ideal lossless capacitive tuning element while preserving the same capacitance value. At $\SI{28}{GHz}$, the incorporation of realistic varactor parasitics introduces approximately $\SI{3.6}{dB}$ of additional reflection loss compared with the ideal lossless capacitive reference ($\SI{1.71}{dB}$). This comparison indicates that the varactor series resistance and associated parasitic loss mechanisms contribute significantly to the unit-cell reflection loss while only marginally affecting the available phase tuning range.

The angular stability of the proposed unit cell is subsequently evaluated for incidence angles ranging from $\theta_i=\SIrange{0}{-45}{\degree}$, as shown in \figref{fig:UC_Norm_vs_oblique}. Although oblique illumination introduces moderate variations in the reflection magnitude, the overall phase-tuning behavior remains well preserved. Moreover, the achievable phase coverage increases from approximately $\SI{300}{\degree}$ under normal incidence to approximately $\SI{315}{\degree}$ at $\theta_i=\SI{-45}{\degree}$. These results confirm the robustness of the proposed unit cell under oblique illumination, which represents the typical operating condition of practical RIS deployments.

The responses presented in \figref{fig:UC_Norm_vs_oblique} highlight the practical limitations imposed by realistic varactor operation, including finite phase availability and reflection-magnitude variations. To assess how these element-level nonidealities influence the radiation performance of the complete RIS, progressively realistic analytical models are introduced in \secref{sec:Analytical_Models}.

\subsection{Waveguide-Based Unit-Cell Measurement and Characterization}\label{sec:UC_Meas}

\begin{figure}[t]
\centering
    \setlength\figurewidth{.45\textwidth}    
    \includegraphics{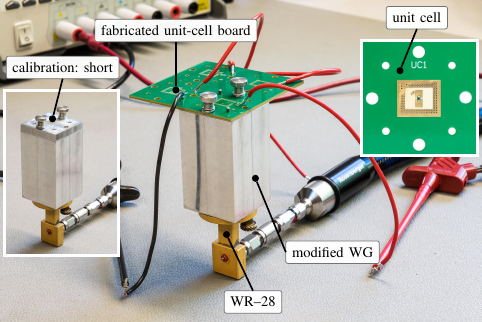}
    \caption{Waveguide-based measurement setup for RIS unit cell characterization, together with the calibration standard and a close-up of the fabricated unit cell.}
  \label{Fig:WG_UC_Meas_Setup}
\end{figure}

To experimentally validate the simulated unit-cell response, the proposed design is characterized using the waveguide-based measurement methodology introduced in~\cite{Manna_APS_2025}. The measurement setup is shown in \figref{Fig:WG_UC_Meas_Setup}, comprising a modified waveguide fixture attached to the flange of a standard WR-28, a calibration standard (short), and the fabricated unit cell under test. Since the measurement concept and parameter-extraction procedure are identical to those reported in~\cite{Manna_APS_2025}, only the essential aspects are summarized here. The adopted calibration procedure enables the extraction of the complex reflection coefficient of the unit cell while minimizing the influence of the measurement fixture.

In contrast to the preliminary investigation reported in~\cite{Manna_APS_2025}, the present study considers a modified unit-cell implementation featuring a different geometry and substrate configuration, requiring the previously extracted unit-cell-dependent varactor model to be reassessed. The unit cells were manufactured using an industrial printed circuit board (PCB) fabrication process and assembled through controlled reflow soldering. Furthermore, four nominally identical unit cells (UC1--UC4) were characterized to assess unit-to-unit repeatability and the consistency of the extracted unit-cell response.

\begin{figure}[ht]
    \centering
    \setlength\figureheight{.52\columnwidth} 
    \setlength\figurewidth{0.72\columnwidth}
    \includegraphics{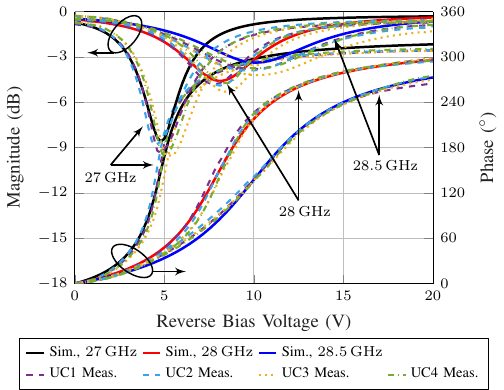}
    \caption{Measured reflection coefficients of four fabricated unit cells (UC1--UC4) together with the corresponding simulated response at $\SI{27}{GHz}$, $\SI{28}{GHz}$, and $\SI{28.5}{GHz}$.}
    \label{fig:UC_meas_27_28_28p5GHz}
    \end{figure}

The measured reflection coefficients of the four fabricated unit cells are compared with the corresponding full-wave simulation results in \figref{fig:UC_meas_27_28_28p5GHz} at $\SI{27}{GHz}$, $\SI{28}{GHz}$, and $\SI{28.5}{GHz}$. Excellent agreement is observed between simulation and measurement for both the reflection magnitude and phase responses across the complete tuning range, with only minor unit-to-unit variations among the four fabricated samples. At the design frequency of $\SI{28}{GHz}$, all four unit cells exhibit phase coverage approaching $\SI{300}{\degree}$, whereas the measured reflection magnitude closely follows the simulated waveguide-embedded response of $\SI{-4.6}{dB}$ with a maximum deviation of less than $\SI{0.7}{dB}$. Similar agreement is observed at $\SI{27}{GHz}$ and $\SI{28.5}{GHz}$, where the measured phase coverage remains within $\SIrange{308}{311}{\degree}$ and approaches $\SI{275}{\degree}$, respectively, while the corresponding simulated reflection magnitudes of $\SI{-8.6}{dB}$ and $\SI{-3.4}{dB}$ are likewise well reproduced. This demonstrates that the extracted model accurately captures both the reflection magnitude and phase characteristics of the proposed unit cell beyond the design frequency.

The observed agreement between simulation and measurement, together with the excellent repeatability across four nominally identical unit cells, validates the extracted unit-cell-dependent varactor model characterized by $L_\mathrm{ser}=\SI{88.5}{pH}$, $C_\mathrm{par}=\SI{9}{fF}$, and $R_\mathrm{ser}=\SI{5.5}{\Omega}$. The inductive and capacitive parasitic parameters $L_\mathrm{ser}$ and $C_\mathrm{par}$ coincide with the independent broadband model reported in~\cite{Escribano_IJEC_2024}, while the extracted series resistance is comparable to the values reported for the same varactor model, ranging from $\SI{7.5}{\Omega}$ around $\SI{21}{GHz}$ in~\cite{Rotshild_IJRFMCAE_2021} to $\SIrange[range-phrase=-]{10.2}{18}{\Omega}$ over a wide frequency range in~\cite{Escribano_IJEC_2024}. 
The value extracted here provides the closest agreement with the measured four unit-cell responses around $\SI{28}{GHz}$.\footnote{The lower effective series resistance $R_\mathrm{ser}$ of the unit-cell embedded varactor diode obtained here, relative to the isolated-device characterization in~\cite{Escribano_IJEC_2024}, is attributed to differences in loss allocation and assembly technique between the two extraction contexts. In the resonant unit cell, part of the dissipation associated with the surrounding metallization and substrate is captured by the periodic structure itself rather than by the lumped resistor of the diode. Additionally, the extraction of series resistance is sensitive to the soldering technique.} Consequently, the experimentally validated unit-cell model serves as the foundation for the analytical modeling and RIS-level performance assessment presented in the subsequent sections. 

\section{RIS Prototype and Analytical Modeling}\label{sec:RIS_Model}
    
\subsection{RIS Architecture and Integrated Control Unit}\label{sec:RIS_Architecture}

\begin{figure}[t]
\centering
\setlength\figurewidth{0.48\columnwidth}
\begin{subfigure}[b]{0.48\columnwidth}
    \centering
    \vspace{0pt}
    \includegraphics[width=\linewidth]{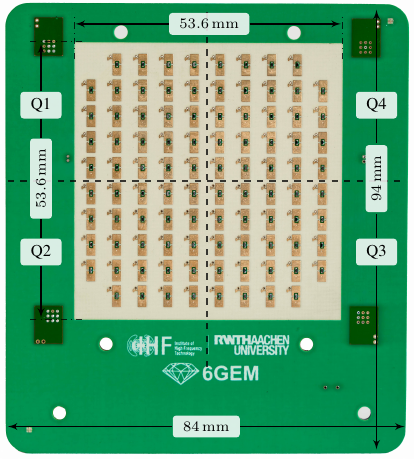}
    \caption{}
    \label{Fig:RIS_Aperture_Top}
\end{subfigure}
\hfill
\begin{subfigure}[b]{0.48\columnwidth}
    \centering
    \vspace{0pt}
    \includegraphics[width=\linewidth]{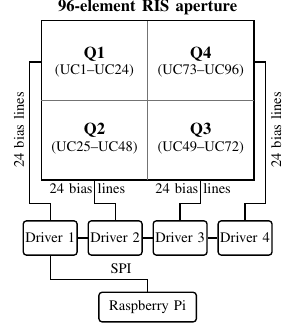}
    \caption{}
    \label{Fig:RIS_system_Architecture}
\end{subfigure}
\caption{Architecture of the proposed $96$-element varactor-based RIS: (a)~fabricated RIS aperture with quadrant-based partitioning (Q1--Q4) and (b)~system-level control architecture for independent bias control of all unit cells.}
\label{Fig:RIS_Architecture}
\end{figure}

The validated unit-cell design was subsequently integrated into a continuously tunable RIS prototype for experimental investigation of programmable wavefront manipulation at $\SI{28}{GHz}$. The fabricated RIS comprises a $10\times10$ unit-cell lattice with four corner elements omitted, resulting in a total of $96$ independently controllable elements. As shown in \figref{Fig:RIS_Aperture_Top}, the active RIS aperture measures $\SI{53.6}{mm}\times\SI{53.6}{mm}$, corresponding to $5\lambda_0\times5\lambda_0$ at $\SI{28}{GHz}$. For implementation and routing purposes, the aperture is partitioned into four quadrants (Q1--Q4), each containing $24$ unit cells. Including the biasing network and integrated control circuitry, the overall prototype dimensions are $\SI{84}{mm}\times\SI{94}{mm}$.


\begin{figure}[t]
    \centering
    \includegraphics[width=\columnwidth]{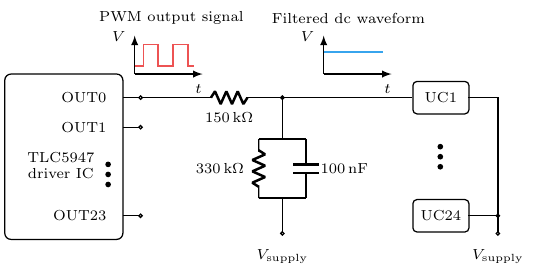}
    \caption{Per-channel PWM-to-dc bias conversion network of one TLC5947 driver, showing the identical low-pass filter applied to each of its $24$ outputs for independent control of the corresponding varactor-loaded unit cells.}
    \label{Fig:pwm_bias_channel}
\end{figure}

The system-level control architecture consists of a Raspberry Pi controller interfaced with four Texas Instruments TLC5947 LED drivers through a serial peripheral interface (SPI), as illustrated in \figref{Fig:RIS_system_Architecture}. The control circuitry is fully integrated on the backside of the RIS PCB, eliminating the need for a separate control board and thereby providing a compact hardware platform. Each TLC5947 provides $24$ independently programmable $12$-bit pulse-width modulation (PWM) channels and is assigned to one RIS quadrant, collectively enabling independent bias control of all $96$ RIS elements. Since the TLC5947 provides PWM outputs whereas the varactors require analog reverse-bias voltages, each bias channel incorporates an identical PWM-to-dc conversion network, as shown in \figref{Fig:pwm_bias_channel}. The network comprises a $\SI{150}{k\Omega}$ series resistor, a $\SI{330}{k\Omega}$ shunt resistor referenced to the supply rail $V_{\mathrm{supply}}=\SI{30}{V}$, and a $\SI{100}{nF}$ capacitor. The resistor pair establishes the effective voltage-divider ratio and contributes to the filtering behavior, while the capacitor provides the primary energy storage for suppressing PWM ripple. Together, these components generate a stable analog reverse-bias voltage for continuous varactor tuning. The $12$-bit PWM resolution provides $4096$ duty-cycle levels, enabling finely resolved analog bias voltages for effective continuous phase tuning. The implemented bias network provides a calibrated reverse-bias voltage range of $\SIrange{0}{20.4}{V}$, compatible with the varactor operating limit.

\begin{figure}[t]
    \centering
    \setlength\figureheight{.20\textwidth}
    \setlength\figurewidth{.40\textwidth}
    \includegraphics{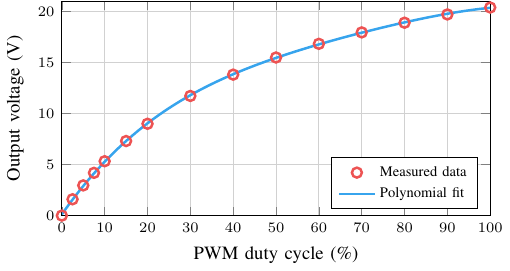}
    \caption{Measured PWM duty-cycle versus output voltage together with the polynomial fitting model used for analog bias-voltage calibration.}
    \label{fig:pwm_voltage_curve}
    \end{figure}

To ensure accurate voltage synthesis across all RIS elements, the PWM-to-dc conversion stage was experimentally calibrated. The measured relationship between the PWM duty-cycle setting and the corresponding output voltage, together with its polynomial approximation, is shown in \figref{fig:pwm_voltage_curve}. The resulting fourth-order calibration model is given by

\begin{equation}
    \label{eq:pwm_cal}
    \begin{aligned}
        V_{\mathrm{out}}(D) ={} & -2.835\times10^{-7}D^4 + 8.017\times10^{-5}D^3 \\
        & -0.009151D^2 + 0.5994D +0.1209,
    \end{aligned}
\end{equation}

where $D$ denotes the PWM duty-cycle percentage and $V_{\mathrm{out}}$ represents the corresponding filtered dc output voltage. The calibration model was integrated into the RIS control framework to enable accurate software-based voltage synthesis and subsequent mapping of the desired phase distributions to the required varactor bias voltages. This enables consistent phase control across the RIS aperture during programmable beamforming operation.

\begin{figure}[t]
\centering
    \setlength\figurewidth{.40\textwidth}    
    \includegraphics{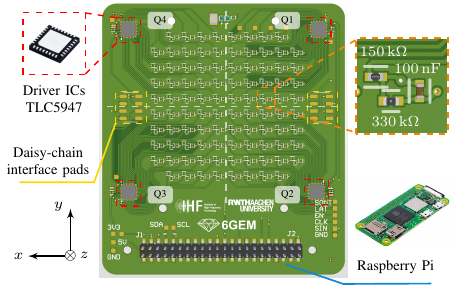}
    \caption{Backside integrated control PCB of the proposed $96$-element varactor-based RIS with distributed bias-generation and scalable digital control circuitry.}
  \label{Fig:backside_pcb}
\end{figure}

The practical implementation of the integrated control platform is shown in \figref{Fig:backside_pcb}. The backside PCB accommodates the driver circuitry, voltage-distribution network, and routing infrastructure required for the $96$ independently controlled bias channels while maintaining compact integration with the RF aperture. To facilitate future array expansion, dedicated edge interconnects are incorporated to support daisy-chaining of multiple RIS modules. Consequently, the proposed control architecture readily scales beyond the presented $96$-element RIS prototype through the same modular hardware concept.

\subsection{Feed-Distance Selection and Aperture-Efficiency Assessment}\label{sec:Feed_Aperture}

The feed distance determines the spatial distribution of the incident field across the RIS aperture and, therefore, influences both the analytical modeling and the experimental validation presented in \secref{sec:Analytical_Models} and \secref{sec:RIS_Meas}, respectively. A first-order estimate of the illuminated region is obtained from the feed half-power beamwidth as
\begin{equation}
D_{\mathrm{fp}} \approx 2r_{\mathrm{TX}}\tan\bigl(\Theta_{\mathrm{HPBW}}/2\bigr),
\end{equation}
where $D_{\mathrm{fp}}$ denotes the illumination footprint and $r_{\mathrm{TX}}$ is the feed-to-RIS distance. For the employed $\SI{28}{GHz}$ horn antenna, the measured E-plane half-power beamwidth of $\Theta_{\mathrm{HPBW}}=\SI{15.5}{\degree}$ results in an illumination footprint of approximately $\SI{54}{mm}$ at $r_{\mathrm{TX}}=\SI{200}{mm}$, closely matching the active RIS aperture and corresponding to an edge taper of $\SI{3}{dB}$.

To quantitatively assess this illumination geometry, the aperture efficiency is evaluated following~\cite{Rahmat_Samii_MOTL_2010} as
\begin{equation}
\eta_{\mathrm{a}} = \eta_{\mathrm{s}} \cdot \eta_{\mathrm{i}},
\end{equation}
where $\eta_{\mathrm{s}}$ denotes the spillover efficiency and $\eta_{\mathrm{i}}$ the illumination efficiency. For the discretized RIS, these quantities are computed as
\begin{equation}
\eta_{\mathrm{s}} =
\frac{\sum\limits_{n} U_n \, \frac{\cos\theta_{\mathrm{inc},n}}{r_n^2} \, \Delta A}
{P_{\mathrm{tot}}},
\end{equation}
and
\begin{subequations}\label{eq:illumination_efficiency}
\begin{align}
\eta_{\mathrm{i}} &=
\frac{\left|\sum\limits_{n} E_n \Delta A \right|^2}
{A_{\mathrm{ap}} \sum\limits_{n} |E_n|^2 \Delta A},
\label{eq:illumination_efficiencya}
\\
\intertext{with the aperture field amplitude}
E_n &= \frac{\sqrt{U_n}\,\cos^{q_e}(\theta_{\mathrm{inc},n})}{r_n}.
\label{eq:illumination_efficiencyb}
\end{align}
\end{subequations}
Here, $U_n$ denotes the feed power pattern evaluated toward the $n$-th unit cell, $r_n$ the distance between the feed and the corresponding element, $\theta_{\mathrm{inc},n}$ the incidence angle, $q_e=1$ the element-pattern exponent, $\Delta A$ the unit-cell area, and $A_{\mathrm{ap}}$ the physical aperture area. The total radiated power over the forward hemisphere is denoted by $P_{\mathrm{tot}}$.

\noindent\textbf{Cosine-Power Model:}
To obtain an initial estimate of the aperture efficiency, the feed is approximated by an ideal symmetric cosine-power pattern following~\cite{Rahmat_Samii_MOTL_2010}. This model predicts $\eta_{\mathrm{s}}\approx46.8\%$, $\eta_{\mathrm{i}}\approx98.5\%$, and $\eta_{\mathrm{a}}\approx46.1\%$ at $r_{\mathrm{TX}}=\SI{200}{mm}$.

\noindent\textbf{Horn-Based Model:}
To account for the actual illumination employed throughout this work, a more realistic feed model is constructed from the measured E- and H-plane horn patterns,
\begin{equation}
    U_{\mathrm{horn}}(\theta_{\mathrm{f}},\phi_{\mathrm{f}}) =
    U_{\mathrm{H}}(\theta_{\mathrm{f}})\cos^2(\phi_{\mathrm{f}}) +
    U_{\mathrm{E}}(\theta_{\mathrm{f}})\sin^2(\phi_{\mathrm{f}}),
\end{equation}
where $(\theta_{\mathrm{f}},\phi_{\mathrm{f}})$ denote the elevation and azimuth angles in the same local feed coordinate system. The corresponding total radiated power is obtained numerically as
\begin{equation}
    P_{\mathrm{tot,horn}}
    =
    \int_{0}^{2\pi}\int_{0}^{\pi/2}
    U_{\mathrm{horn}}(\theta_{\mathrm{f}},\phi_{\mathrm{f}})
    \sin\theta_{\mathrm{f}}\,
    \mathrm{d}\theta_{\mathrm{f}}\,
    \mathrm{d}\phi_{\mathrm{f}}.
\end{equation}
Using this measured horn model yields $\eta_{\mathrm{s}}\approx31.8\%$, $\eta_{\mathrm{i}}\approx98.7\%$, and $\eta_{\mathrm{a}}\approx31.4\%$ at $r_{\mathrm{TX}}=\SI{200}{mm}$.

For both illumination models, the illumination efficiency remains close to unity, indicating that the incident field amplitude is nearly uniform across the finite RIS aperture. Consequently, the overall aperture efficiency is primarily limited by spillover rather than illumination taper. Although reducing the feed distance to $r_{\mathrm{TX}}=\SI{100}{mm}$ increases the theoretical aperture efficiency ($\eta_{\mathrm{a}}\approx56.3\%$), the resulting geometry is incompatible with the employed compact antenna test range (CATR) measurement configuration due to blockage and shadowing effects (see \secref{sec:NF_FF}). Therefore, a feed distance of $r_{\mathrm{TX}}=\SI{200}{mm}$ is adopted as a practical compromise between aperture efficiency and measurement feasibility.

\subsection{Analytical Modeling of RIS Beam Steering with Continuous Phase Control}\label{sec:Analytical_Models}

The electromagnetic response of the RIS is fundamentally governed by the complex field distribution established across its aperture, which results from the interaction between the incident feed illumination and the local reflection characteristics of the individual unit cells. The proposed analytical framework progressively incorporates realistic electromagnetic effects, from practical horn-feed illumination to unit-cell nonidealities, allowing their individual and combined impacts on RIS beamforming performance to be systematically isolated and quantified. The resulting reflected aperture field at the \mbox{$n$-th} RIS element is expressed as
\begin{equation}
    E_n^{\textrm{RIS}} = |E_{\textrm{TX},n}| |\Gamma_n| e^{-j(\Phi_n - \phi_{\textrm{inc},n})},
    \label{eq:Aperture_field_basic}
\end{equation}
where $|E_{\mathrm{TX},n}|$ denotes the incident field amplitude across the RIS aperture, $|\Gamma_n|$ and $\Phi_n$ represent the magnitude and phase of the unit-cell reflection coefficient, respectively, and $\phi_{\mathrm{inc},n}$ is the geometric phase delay from the feed antenna to the \mbox{$n$-th} RIS element. This unified aperture-field formulation forms the basis for all subsequent analytical models. To account for the practical feed illumination, the incident field is modeled following~\cite{Encinar_reflectarray_2007} as
\begin{equation}
    E_{\textrm{TX},n}^{\textrm{horn}} = \frac{F(\theta_n,\phi_n)}{R_n} e^{-jkR_n},
\end{equation}
where $F(\theta_n,\phi_n)$ denotes the measured horn radiation pattern evaluated in the direction of the $n$-th RIS element, and $R_n$ is the corresponding feed-to-element distance. Unlike conventional analytical formulations based on isotropic point-source illumination, the proposed framework employs the measured E- and H-plane horn radiation patterns. This illumination model incorporates both spherical-wave propagation and realistic feed taper, enabling accurate prediction of aperture illumination, sidelobe behavior, and beamforming performance.

\begin{figure}[t]
\centering
\subcaptionbox{\hspace*{19.5pt}\label{Fig:Phase_analytical}}[0.5\columnwidth][c]{%
    \includegraphics{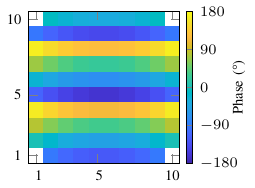}}%
\subcaptionbox{\hspace*{19.5pt}\label{Fig:Phase_CST}}[0.5\columnwidth][c]{%
    \includegraphics{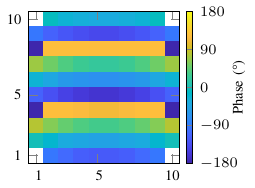}}\\[0pt]
\subcaptionbox{\hspace*{19.5pt}\label{Fig:Phase_difference}}[0.5\columnwidth][c]{%
    \includegraphics{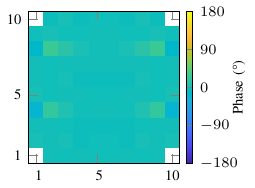}}%
\subcaptionbox{\hspace*{19.5pt}\label{Fig:Mag_CST}}[0.5\columnwidth][c]{%
    \includegraphics{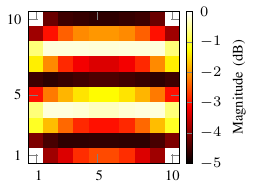}}%
\caption{Element-wise phase and magnitude distributions of the proposed RIS for beam steering toward $\theta_\textrm{RX}=\SI{0}{\degree}$ in the $\phi=\SI{90}{\degree}$ plane, illuminated from $\theta_\textrm{TX}=\SI{-30}{\degree}$ in the same azimuthal plane and $r_\textrm{TX}=\SI{200}{mm}$. Shown are (a)~the ideal continuous reflection phase, (b)~the CST-mapped reflection phase obtained using the closest available unit-cell state, (c)~the resulting phase error, and (d)~the corresponding unit-cell reflection magnitude distribution.}
\label{Fig:Phase_Mag_variations}
\end{figure}

Based on the unified aperture-field formulation, the reflected field at each RIS element is governed by the phase difference $(\Phi_n-\phi_{\mathrm{inc},n})$, which determines the aperture phase distribution and consequently the far-field radiation pattern. For a desired beam-steering direction, a target reflection phase $\Phi_{\mathrm{req},n}$ is synthesized from the geometrical path difference associated with the feed antenna, the RIS element, and the observation direction. In an ideal continuously tunable RIS, this phase distribution can be realized exactly. In practice, however, the achievable reflection phase is constrained by the unit-cell characteristics, resulting in a limited realized phase $\Phi_{\mathrm{sim},n}$ obtained from full-wave simulations. For the representative beam-steering configuration, the corresponding required ideal phase, realized phase, and resulting phase-error distributions are illustrated in \figref{Fig:Phase_Mag_variations}(a)--(c), respectively.

\noindent\textbf{Progressive Nonideality Models:}
Building upon the unified aperture-field formulation, four progressively realistic analytical models are established to systematically isolate the individual contributions of unit-cell loss and limited phase availability to the overall beamforming performance.

\textit{Model I -- Ideal Phase Synthesis Model}:
This model assumes lossless unit cells ($|\Gamma_n|=1$) with ideal continuous phase control, allowing the synthesized phase $\Phi_{\mathrm{req},n}$ to be realized exactly. The corresponding aperture field is given by
\begin{equation}
    E_n^{\textrm{RIS,I}} = |E_{\textrm{TX},n}| e^{-j\left(\Phi_{\textrm{req},n} - \phi_{\textrm{inc},n}\right)}.
\end{equation}
This model establishes the theoretical upper performance limit of a continuously tunable RIS.

\textit{Model II -- Loss-Aware Ideal Phase Model}:
This model incorporates the simulated unit-cell magnitude response while retaining ideal phase synthesis with full $\SI{360}{\degree}$ phase availability. The corresponding aperture field becomes
\begin{equation}
    E_n^{\textrm{RIS,II}} = |E_{\textrm{TX},n}| |\Gamma_{\textrm{sim},n}| e^{-j\left(\Phi_{\textrm{req},n} - \phi_{\textrm{inc},n}\right)}.
\end{equation}
The resulting unit-cell magnitude distribution for the representative beam-steering configuration is illustrated in \figref{Fig:Mag_CST}. This model isolates the impact of practical reflection losses while preserving ideal phase control.

\textit{Model III -- Phase-Realistic Model}:
This model assumes lossless reflection but replaces the ideal phase with the simulated unit-cell phase response obtained from full-wave analysis,
\begin{equation}
    E_n^{\textrm{RIS,III}} = |E_{\textrm{TX},n}| e^{-j\left(\Phi_{\textrm{sim},n} - \phi_{\textrm{inc},n}\right)}.
\end{equation}
The corresponding phase error is defined as $\Delta\Phi_n=\Phi_{\mathrm{req},n}-\Phi_{\mathrm{sim},n}$. The resulting phase inaccuracies are visualized in \figref{Fig:Phase_difference}, allowing the degradation introduced by the limited phase tuning range to be quantified independently of magnitude loss.

\textit{Model IV -- Full-Wave Consistent Model}:
Finally, both the simulated magnitude and phase responses are incorporated,
\begin{equation}
    E_n^{\textrm{RIS,IV}} = |E_{\textrm{TX},n}| |\Gamma_{\textrm{sim},n}| e^{-j\left(\Phi_{\textrm{sim},n} - \phi_{\textrm{inc},n}\right)},
    \label{eq:RIS_Field_Model_IV}
\end{equation}
thereby providing the closest analytical approximation to the experimentally realizable RIS response.

The aperture fields defined by Models I--IV are transformed into far-field radiation patterns using the common array-factor formulation
\begin{subequations}\label{eq:AF}
\begin{align}
\mathrm{AF}(\theta,\phi)
&=
\sum_{n=1}^{N}
w_n
\,e^{-jk\,\hat{\mathbf{r}}(\theta,\phi)\cdot\mathbf{r}_n},
\label{eq:AFa}
\\
\intertext{where the complex aperture weights are defined as}
w_n
&=
E_n^{\mathrm{RIS}}.
\label{eq:AFb}
\end{align}
\end{subequations}
Substituting the corresponding aperture-field expressions of Models I--IV into $w_n$ directly yields the associated radiation patterns. For consistent comparison with the experimental measurements, all analytical results are normalized with respect to the specular reflection from a reference metallic plate of identical dimensions. Since the metal plate preserves the incident horn illumination while exhibiting a spatially uniform aperture phase, it provides a physically meaningful reference independent of the applied RIS phase distribution. Accordingly, the normalized radiation patterns are obtained as
\begin{equation}
    \textrm{AF}_{\textrm{norm}}(\theta,\phi) = \frac{\textrm{AF}(\theta,\phi)}{\textrm{AF}_{\textrm{metal,max}}},
\end{equation}
where $\mathrm{AF}_{\mathrm{metal,max}}$ denotes the maximum magnitude of the metallic reference response.

\begin{figure}[t]
\centering
    \setlength\figureheight{.25\textwidth}
    \setlength\figurewidth{.40\textwidth}
    \includegraphics{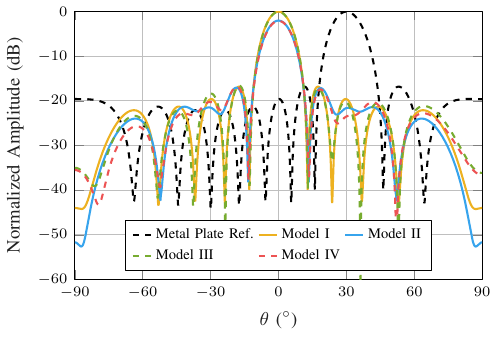}
    \caption{Comparison of the radiation patterns predicted by the four progressively realistic analytical models of the proposed RIS at $\SI{28}{GHz}$, corresponding to the beam-steering configuration of \figref{Fig:Phase_Mag_variations}.}
  \label{Fig:Beamsteering_Analytical}
\end{figure}

The normalized radiation patterns obtained from Models I--IV together with the metallic reference are presented in \figref{Fig:Beamsteering_Analytical}. The progression from Model I to Model IV systematically incorporates the unit-cell nonidealities, isolating their individual and combined effects on RIS beamforming performance. Model I represents the ideal baseline; Model II isolates the effect of unit-cell loss, Model III quantifies the impact of limited phase availability, and Model IV captures their combined influence. The progressive deviation observed across the four models demonstrates the cumulative degradation introduced by these nonidealities, whereas the metallic reference establishes the baseline specular reflection of the illuminated aperture. Collectively, the proposed analytical framework bridges the gap between ideal analytical beamforming and experimentally realizable RIS performance.

\section{Experimental Validation of Beam Steering and Programmable Reconfigurability}\label{sec:RIS_Meas}

\subsection{Near-Field-to-Near-Field Characterization}\label{sec:NF_NF}
\begin{figure}[ht]
\centering
    \setlength\figurewidth{.49\textwidth}    
    \includegraphics{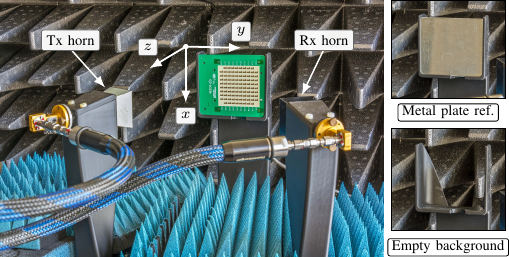}
    \caption{Near-field-to-near-field characterization setup in the $\phi=\SI{90}{\degree}$ plane ($yz$-plane) with $\theta_{\mathrm{TX}}=\SI{-30}{\degree}$ and $\theta_{\mathrm{RX}}=\SI{30}{\degree}$ at a Tx--RIS--Rx separation of $\SI{200}{mm}$, together with the metal-plate and empty-background references.}
  \label{Fig:NF_to_NF_Meas_Setup}
\end{figure}

\begin{figure}[t]
  \centering
  \setlength\figureheight{0.26\columnwidth}
  \setlength\figurewidth{0.33\columnwidth}
  %
  \subcaptionbox{\hspace*{0.45cm}\label{Fig:2D_Mag_response}}[0.49\columnwidth][c]{%
    \includegraphics[scale=0.995]{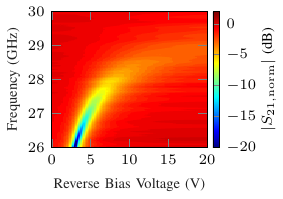}}%
  \hfill
  \subcaptionbox{\hspace*{0.45cm}\label{Fig:2D_Phase_response}}[0.49\columnwidth][c]{%
    \includegraphics[scale=0.995]{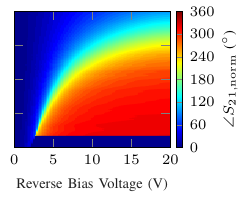}}%
  \caption{Measured near-field-to-near-field characterization of the normalized
  transmission response of the RIS: (a)~magnitude and (b)~phase.}
  \label{Fig:2D_Mag_&_Phase_response}
\end{figure}

The experimentally validated unit-cell response reported in \secref{sec:UC_Meas} is next reassessed after integration into the complete RIS, where the measured response inherently includes practical array effects such as inter-element coupling. The characterization is performed using the bistatic near-field-to-near-field setup shown in \figref{Fig:NF_to_NF_Meas_Setup}, in which two standard horn antennas are fixed at a Tx--RIS--Rx separation of $\SI{200}{mm}$ and oriented at $\theta_{\mathrm{TX}}=\SI{-30}{\degree}$ and $\theta_{\mathrm{RX}}=\SI{30}{\degree}$ in the $\phi=\SI{90}{\degree}$ plane, corresponding to the E-plane. During the measurement, all 96 unit cells are simultaneously biased with a common reverse-bias voltage swept from $\SIrange{0}{20}{V}$. To suppress background contributions and reference the response to an equal-sized metallic reflector, the measured transmission coefficient is normalized according to
\begin{equation}
S_{21,\mathrm{norm}}=\frac{S_{21,\mathrm{RIS}}-S_{21,\mathrm{bg}}}{S_{21,\mathrm{metal}}-S_{21,\mathrm{bg}}},
\end{equation}
where $S_{21,\mathrm{RIS}}$, $S_{21,\mathrm{metal}}$, and $S_{21,\mathrm{bg}}$ denote the measured transmission coefficients of the RIS, the metallic reference, and the empty-background measurement, respectively. The resulting normalized magnitude and phase responses are presented in \figref{Fig:2D_Mag_&_Phase_response}.

Unlike the discrete frequency comparisons presented in \figref{fig:UC_meas_27_28_28p5GHz}, the measured response is characterized over the $\SIrange{26}{30}{GHz}$ frequency range, providing a comprehensive experimental view of the programmable behavior of the complete RIS. The measured responses remain consistent with the unit-cell characterization in \secref{sec:UC_Meas}, exhibiting approximately $\SI{300}{\degree}$ phase coverage with a magnitude of $\SI{-5.41}{dB}$ at $\SI{28}{GHz}$. Similar agreement is observed at the adjacent frequencies, achieving $\SI{315.2}{\degree}$ and $\SI{271.2}{\degree}$ phase coverage with corresponding reflection magnitudes of $\SI{-9.3}{dB}$ and $\SI{-3.83}{dB}$ at $\SI{27}{GHz}$ and $\SI{28.5}{GHz}$, respectively. These results confirm that the experimentally validated unit-cell response is preserved after integration into the complete RIS aperture, thereby establishing the experimental link between the isolated unit-cell characterization of \secref{sec:UC_Meas} and the array-level beam-steering validation presented in the following subsection.
   
\subsection{Near-Field-to-Far-Field Beam-Steering Validation}\label{sec:NF_FF}

\begin{figure*}[!t]
\centering
    \includegraphics[width=\textwidth]{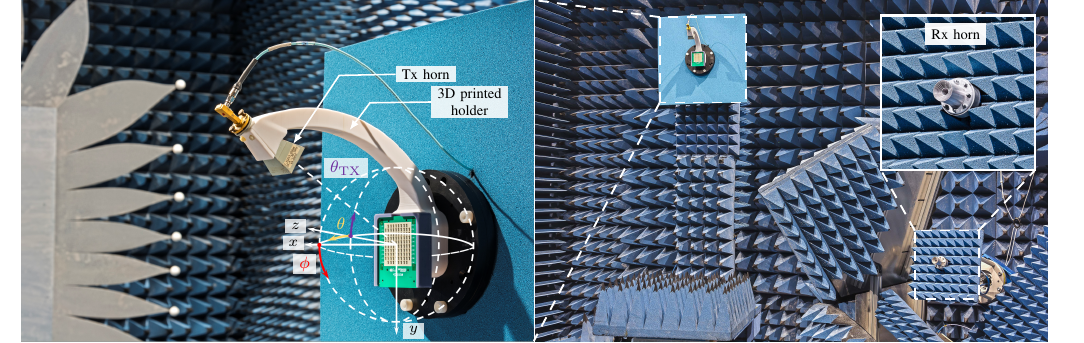}
    \caption{Near-field-to-far-field measurement setup in the compact antenna test range (CATR). The RIS is illuminated by a standard \SI{28}{GHz} horn at $r_{\mathrm{TX}}=\SI{200}{mm}$ with $\theta_{\mathrm{TX}}=\SI{-30}{\degree}$ and $\phi_{\mathrm{TX}}=\SI{90}{\degree}$ ($yz$-plane excitation). Insets show the overall CATR arrangement and the corrugated receiving horn used for far-field measurements.}
  \label{Fig:CATR_RIS_Meas_Setup}
\end{figure*}

The beam-steering performance of the fabricated RIS is experimentally validated using the compact antenna test range (CATR) setup shown in \figref{Fig:CATR_RIS_Meas_Setup}. The RIS is illuminated by a standard $\SI{28}{GHz}$ linearly polarized horn antenna positioned within the radiating near-field region of the RIS at $r_{\mathrm{TX}}=\SI{200}{mm}$ with $\theta_{\mathrm{TX}}=\SI{-30}{\degree}$ and $\phi_{\mathrm{TX}}=\SI{90}{\degree}$, while the reflected far-field radiation patterns are measured using a corrugated circular receiving horn.\footnote{In the CATR setup, the corrugated horn serves as the feed antenna for the CATR's reflector. For consistency with analytical framework presented in \secref{sec:Feed_Aperture}, the transmit (Tx) and receive (Rx) designations are interchanged relative to the actual measurement. This interchange has no effect on the measured results due to reciprocity.}
The chosen illumination geometry follows the aperture-efficiency assessment presented in \secref{sec:Feed_Aperture}. A dedicated 3D-printed fixture is used to accurately position and align both the RIS and the transmitting horn. This arrangement permits measurements over the full spherical angular range while introducing only minimal shadowing, thereby supporting reliable comparison with the corresponding simulations~\cite{Reher_AMTA_2023,Manna_EuMC_2026}. All radiation patterns are normalized with respect to the maximum specular reflection of an equal-sized metallic reference, thereby enabling consistent comparison among analytical predictions, full-wave simulations, and experimental measurements.

\begin{figure}[t] 
\centering
    \begin{subfigure}[c]{\columnwidth}
        \centering
        \includegraphics{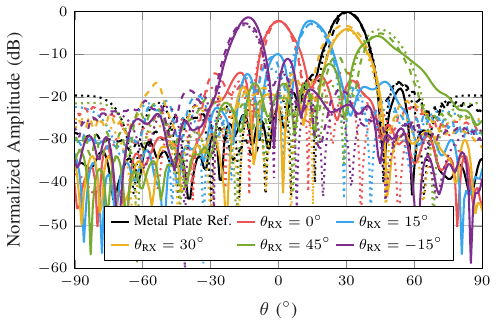}
        \label{Fig:Meas_90}
    \end{subfigure}

    \begin{subfigure}[c]{\columnwidth}
        \centering
        \includegraphics{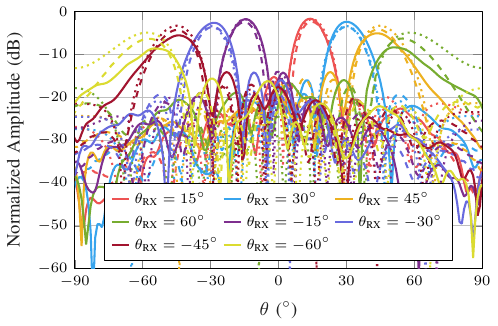}
        \label{Fig:Meas_0}
    \end{subfigure}

    \begin{subfigure}[c]{\columnwidth}
        \centering
        \includegraphics{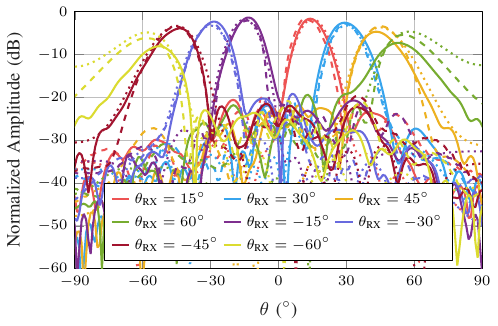}
        \label{Fig:Meas_45}
    \end{subfigure}

    \caption{Analytical, simulated, and measured radiation patterns at \SI{28}{GHz} for beam steering in the $\phi=\SI{90}{\degree}$ (top), $\phi=\SI{0}{\degree}$ (middle), and $\phi=\SI{45}{\degree}$ (bottom) planes. All patterns are normalized to the maximum specular reflection of an equal-sized metallic reflector. Dotted, dashed, and solid lines denote the analytical, simulated, and measured results, respectively.}
  \label{Fig:Beamsteering_plots}
\end{figure}

The measured beam-steering performance is presented in \figref{Fig:Beamsteering_plots}, where the top row shows the principal $\phi=\SI{90}{\degree}$ plane, while the middle and bottom rows present additional validations in the $\phi=\SI{0}{\degree}$ and $\phi=\SI{45}{\degree}$ planes. In the principal plane, excellent agreement is observed among the analytical predictions obtained using Model IV, full-wave simulations, and experimental measurements for steering angles from $\theta_{\mathrm{RX}}=\SI{-15}{\degree}$ to $\SI{45}{\degree}$. The close agreement confirms that the experimentally characterized unit-cell response and the proposed analytical beamforming framework accurately predict the measured far-field beam-steering behavior of the fabricated RIS. Owing to the illumination geometry, angular regions in the vicinity of the transmitting horn are intentionally excluded from the measurements because of unavoidable physical blockage. For the largest steering angle of $\theta_{\mathrm{RX}}=\SI{45}{\degree}$ in the principal plane, the measured beam peak occurs at approximately $\SI{43}{\degree}$, corresponding to a deviation of $\SI{-2}{\degree}$, together with a measured scan loss of approximately $\SI{3.6}{dB}$. Excellent agreement is likewise maintained in the $\phi=\SI{0}{\degree}$ and $\phi=\SI{45}{\degree}$ planes up to
$\theta_{\mathrm{RX}}=\pm\SI{45}{\degree}$, with the beam-peak deviation remaining within $\SI{2}{\degree}$. Larger deviations appear only for the most extreme steering angles of
$\theta_{\mathrm{RX}}=\pm\SI{60}{\degree}$. These discrepancies are primarily attributed to the finite electrical aperture of the compact $5\lambda_0\times5\lambda_0$ RIS together with the reduced practical phase availability discussed in \secref{sec:Feed_Aperture}. Minor deviations between simulation and measurement may additionally arise from fabrication tolerances, including the slight PCB curvature observed after fabrication due to asymmetric PCB stackup. In the $\phi=\SI{90}{\degree}$ and $\phi=\SI{0}{\degree}$ planes, the measured sidelobe levels remain within approximately $\SIrange{-9}{-13}{dB}$, whereas the $\phi=\SI{45}{\degree}$ plane exhibits improved sidelobe suppression of $\SIrange{-18}{-22}{dB}$.


\begin{figure}[ht]
\centering
    \setlength\figureheight{.24\textwidth}
    \setlength\figurewidth{.40\textwidth}
    \includegraphics{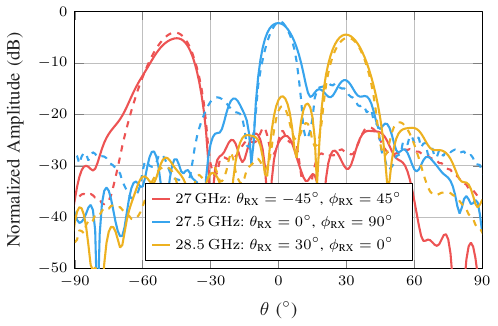}
    \caption{Radiation patterns of the proposed RIS at $\SI{27}{GHz}$, $\SI{27.5}{GHz}$, and $\SI{28.5}{GHz}$ for representative steering angles and observation planes. Dashed and solid lines denote the simulated and measured results, respectively.}
  \label{Fig:Beamsteering_RIS_diff_freq}
\end{figure}

The beam-steering performance of the fabricated RIS beyond the design frequency is further evaluated in \figref{Fig:Beamsteering_RIS_diff_freq} using representative steering configurations at $\SI{27}{GHz}$, $\SI{27.5}{GHz}$, and $\SI{28.5}{GHz}$. Despite the frequency-dependent variation of the unit-cell response discussed in \secref{sec:UC_Meas}, the shown cases span different steering angles and observation planes while maintaining good agreement between simulations and measurements. These results further confirm that the fabricated RIS preserves effective beam-steering capability over the considered operating bandwidth of $\SI{5.41}{\percent}$.\footnote{The fractional bandwidth is defined over the experimentally evaluated span of \SIrange{27}{28.5}{GHz} relative to its center frequency \SI{27.75}{GHz}.}

Collectively, the close agreement among the analytical predictions, full-wave simulations, and experimental measurements validates the proposed beam-steering framework and experimentally confirms the capability of the fabricated RIS to realize accurately controlled continuous-phase wavefront synthesis.

\subsection{Phase Quantization Using the Continuous RIS Platform}\label{sec:Quantization}

The continuously tunable RIS presented in this work also provides a versatile experimental platform for investigating finite-bit phase quantization using the same fabricated hardware. Unlike conventional studies that require separate RIS implementations for different phase resolutions~\cite{Rocca_TAP_2022,Sravan_TAP_Part1_2025,Shamim_TAP_2025}, the proposed platform enables direct experimental emulation of 3-bit, 2-bit, and 1-bit phase control by applying the corresponding quantized phase distributions to the continuously tunable aperture. The experimentally validated continuous phase coverage (e.g., $\SI{300}{\degree}$ at $\SI{28}{GHz}$) is uniformly partitioned into 8, 4, and 2 discrete phase states to emulate 3-bit, 2-bit, and 1-bit phase quantization, respectively.

    \begin{figure}[t]
    \centering
    \setlength\figureheight{.25\textwidth}
    \setlength\figurewidth{.40\textwidth}
    \includegraphics{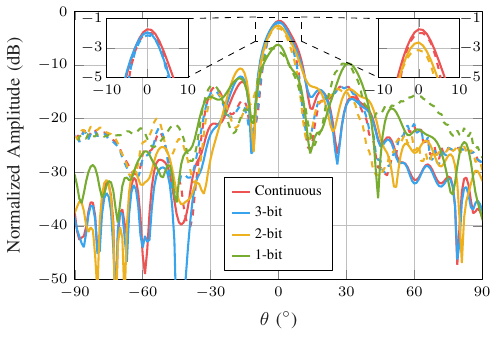}
    \caption{Comparison of continuous and quantized phase control schemes for a selected beam-steering configuration in the principal plane at $\SI{28}{GHz}$. Dashed and solid lines denote the simulated and measured results, respectively.}
    \label{fig:Quant_beamsteering}
    \end{figure}

The experimental emulation of finite-bit phase control is validated in \figref{fig:Quant_beamsteering}, where the continuous-phase reference is compared with its 3-bit, 2-bit, and 1-bit counterparts for a representative beam-steering configuration from $\theta_{\mathrm{TX}}=\SI{-30}{\degree}$, $\phi_{\mathrm{TX}}=\SI{90}{\degree}$ to $\theta_{\mathrm{RX}}=\SI{0}{\degree}$, $\phi_{\mathrm{RX}}=\SI{90}{\degree}$. Good agreement between simulations and measurements is maintained for all quantization levels. As expected, the beamforming performance progressively degrades with decreasing phase resolution, where the 3-bit implementation remains nearly indistinguishable from the continuous-phase reference. The 2-bit implementation introduces only a modest degradation, and the 1-bit implementation exhibits the largest quantization loss together with the most pronounced quantization lobes. A similar trend is observed for the sidelobe levels, where the 3-bit and 2-bit implementations remain close to the continuous-phase case, whereas the binary implementation exhibits noticeably increased sidelobes, particularly in the principal plane.

\begin{figure}[t]
    \centering
    \setlength\figureheight{.20\textwidth}
    \setlength\figurewidth{.40\textwidth}
    \includegraphics{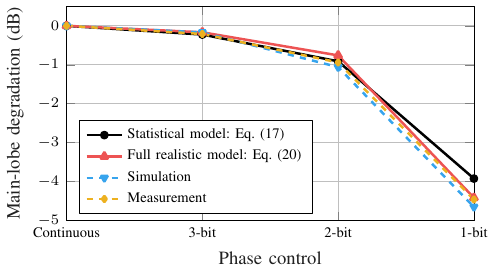}
    \caption{Quantization-induced main-lobe degradation relative to continuous phase control, comparing statistical, full realistic, simulated, and measured results.}
    \label{fig:Quant_loss_1D}
    \end{figure}

To quantitatively predict the experimentally observed quantization degradation, a theoretical baseline is first established using the classical statistical approximation. For an ideal uniformly distributed $L=2^b$-level phase quantizer, where $b$ denotes the number of quantization bits, the corresponding power-domain quantization efficiency is given by~\cite{Romanofsky_TAP_2005}
\begin{equation}
\eta_{\mathrm{stat}}
=
\left(
\frac{\sin(\pi/L)}
{\pi/L}
\right)^2,
\label{eq:Quant_eff_Stat_model}
\end{equation}
for which the corresponding statistical quantization loss is obtained as $\Delta L_{\mathrm{stat}}=10\log_{10}\left(\eta_{\mathrm{stat}}\right)$. Although this classical expression provides a useful theoretical estimate, it neglects the realistic horn illumination, state-dependent unit-cell magnitude, and actual quantized phase distribution. The unified aperture-field formulation of \secref{sec:Analytical_Models}, particularly \eqref{eq:RIS_Field_Model_IV} corresponding to Model~IV, is therefore extended to finite-bit operation. For a $b$-bit RIS, the reflected field of the $n$-th element is written as
\begin{equation}
E_n^{\mathrm{RIS},Q}
=
|E_{\mathrm{TX},n}|
|\Gamma_{Q,n}|
e^{-j\left(\Phi_{Q,n}-\phi_{\mathrm{inc},n}\right)},
\end{equation}
where $\Phi_{Q,n}$ and $|\Gamma_{Q,n}|$ are the selected quantized phase state and its corresponding reflection magnitude, respectively. The resulting coherence efficiency is
\begin{equation}
\eta_Q
=
\frac{
\left|
\displaystyle\sum_{n=1}^{N}
|E_{\mathrm{TX},n}|\,|\Gamma_{Q,n}|\,
e^{j\Delta\Phi_{Q,n}}
\right|^{2}
}
{
\left(
\displaystyle\sum_{n=1}^{N}
|E_{\mathrm{TX},n}|\,|\Gamma_{Q,n}|
\right)^{2}
},
\end{equation}
where $\Delta\Phi_{Q,n}=\Phi_{\mathrm{req},n}-\Phi_{Q,n}$ is the residual phase error. This definition quantifies the loss of coherent aperture addition under the realistic horn illumination and state-dependent unit-cell response. Since a small residual phase error also remains for continuous tuning because of the finite available phase coverage, the same expression is evaluated using $\Phi_{\mathrm{sim},n}$ and $|\Gamma_{\mathrm{sim},n}|$ to obtain the continuous-phase reference efficiency $\eta_C$. The quantization loss is then defined as
\begin{equation}
\Delta L_Q
=
10\log_{10}\left(\frac{\eta_Q}{\eta_C}\right),
\label{eq:Quant_loss_realistic_model}
\end{equation}
thereby quantifying the degradation of the finite-bit implementation relative to the continuous-phase realization. The resulting comparison is presented in \figref{fig:Quant_loss_1D}, where the classical statistical model provides a useful first-order prediction of the measured degradation, while the proposed fully realistic model accurately predicts the simulated and measured quantization losses, particularly at 1-bit resolution, where the statistical estimate under-predicts the measured loss by $\SI{0.54}{dB}$.

\begin{figure}[t]
  \centering
  \setlength\figureheight{0.45\columnwidth}
  \setlength\figurewidth{0.22\columnwidth}
  %
  \subcaptionbox{\hspace*{-32.5pt}\label{Fig:Quant_loss_3bit}}{%
    \includegraphics{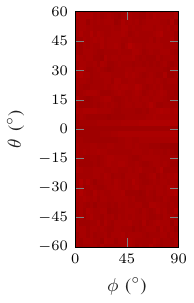}}\hspace{-6pt}%
  \subcaptionbox{\hspace*{0pt}\label{Fig:Quant_loss_2bit}}{%
    \includegraphics{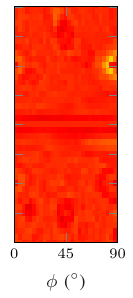}}\hspace{-6pt}%
  \subcaptionbox{\hspace*{31.4pt}\label{Fig:Quant_loss_1bit}}{%
    \includegraphics{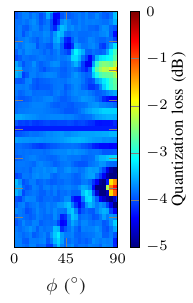}}%
  \caption{Scan-dependent quantization loss relative to the continuous-phase
  reference over the steering range $\theta \in [\SI{-60}{\degree},\SI{60}{\degree}]$
  and $\phi \in [\SI{0}{\degree},\SI{90}{\degree}]$ for (a)~3-bit, (b)~2-bit, and
  (c)~1-bit phase control.}
  \label{Fig:Quant_loss_2D}
\end{figure}

Since the proposed fully realistic model accurately predicts the measured quantization loss, it is subsequently employed to analytically evaluate the scan-dependent quantization degradation over the complete steering region ($\theta\in[\SI{-60}{\degree},\SI{60}{\degree}]$ and $\phi\in[\SI{0}{\degree},\SI{90}{\degree}]$), as presented in \figref{Fig:Quant_loss_2D}. The results show that the quantization loss depends on both the steering angle and the observation plane, while 3-bit operation remains nearly indistinguishable from continuous phase control over most of the investigated scan region, consistent with~\cite{Yang_AWPL_2017,Sravan_TAP_Part1_2025}. The two low-loss regions in
\figref{Fig:Quant_loss_1bit} correspond to binary-compatible phase distributions: the
specular direction ($\theta_{\mathrm{RX}}=\SI{30}{\degree}$), where the required
inter-element phase progression vanishes, and
$\theta_{\mathrm{RX}}=\SI{-30}{\degree}$, where it reaches $\SI{180}{\degree}$. Collectively, these results demonstrate that the proposed continuously tunable RIS provides not only an experimentally validated beam-steering platform but also a versatile experimental testbed for investigating finite-bit RIS architectures and their associated quantization losses.


\subsection{Far-Field-to-Far-Field Angular Response Assessment}\label{sec:FF_FF}

\begin{figure}[t]
\centering
    \setlength\figurewidth{.45\textwidth}    
    \includegraphics{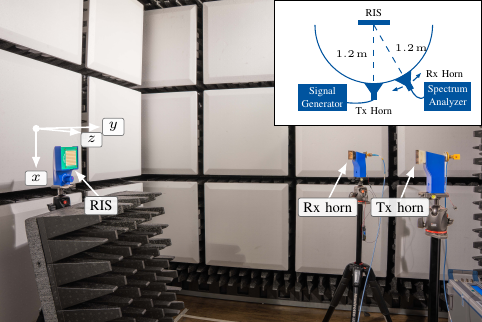}
    \caption{Far-field-to-far-field angular-response measurement setup in the $\phi=\SI{90}{\degree}$ plane with fixed $\theta_{\mathrm{TX}}=\SI{0}{\degree}$, while the receiving horn is scanned over $\theta_{\mathrm{RX}}\in[\SI{-50}{\degree},\SI{50}{\degree}]$. The Tx--RIS and RIS--Rx separations are $\SI{1.2}{m}$.}
  \label{Fig:FF_to_FF_Meas_Setup}
\end{figure}

To further validate the beam-steering capability under practical wireless-link conditions, the proposed RIS is evaluated using the far-field-to-far-field configuration shown in \figref{Fig:FF_to_FF_Meas_Setup}. According to the far-field criterion $R_{\mathrm{FF}}=2D_{\mathrm{RIS}}^2/\lambda$, where $D_{\mathrm{RIS}}$ denotes the RIS aperture diagonal, the corresponding far-field distance is approximately $\SI{1.07}{m}$ at $\SI{28}{GHz}$.\footnote{Here $D_{\mathrm{RIS}}$ is taken as the diagonal of the active RIS aperture, $D_{\mathrm{RIS}}=\sqrt{2}\times\SI{53.6}{mm}\approx\SI{75.8}{mm}$ [\figref{Fig:RIS_Aperture_Top}]. Substituting into $R_{\mathrm{FF}}$ at $\SI{28}{GHz}$ yields $\SI{1.07}{m}$, which satisfies the far-field condition for the measurement used in this section.} Therefore, both the transmitting and receiving horn antennas are positioned in the far field of the RIS, with Tx--RIS and RIS--Rx separations of $\SI{1.2}{m}$. The transmitting horn is fixed at $\theta_{\mathrm{TX}}=\SI{0}{\degree}$ and $\phi_{\mathrm{TX}}=\SI{90}{\degree}$, whereas the receiving horn is manually positioned from $\theta_{\mathrm{RX}}=\SI{-50}{\degree}$ to $\SI{50}{\degree}$ in $\SI{10}{\degree}$ increments. A signal generator and spectrum analyzer are employed to directly measure the received over-the-air power at the operating frequency, thereby providing a practical assessment of the RIS-assisted wireless link.

\begin{figure}[t]
\centering

\begin{subfigure}[t]{\columnwidth}
    \centering
    \includegraphics{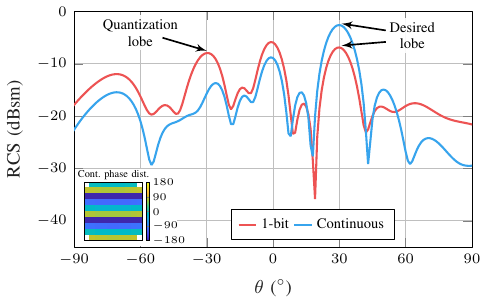}
    \caption{\hspace*{-31.3pt}}
    \label{fig:FF_to_FF_Simulation}
\end{subfigure}
\begin{subfigure}[t]{\columnwidth}
    \centering
    \includegraphics{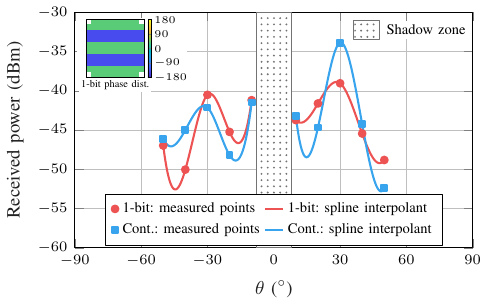}
    \caption{\hspace*{-31.3pt}}
    \label{fig:FF_to_FF_Measurement}
\end{subfigure}
\caption{Comparison of continuous and 1-bit phase control for beam steering at $\SI{28}{GHz}$: (a)~simulated far-field RCS response and (b)~measured far-field-to-far-field angular response. Spline interpolants in (b)~are shown for visual guidance only; the corresponding required continuous and 1-bit phase distributions are included as insets.}
\label{Fig:FF_to_FF_Sim_Meas}
\end{figure}

The corresponding simulated bistatic radar cross section (RCS) and measured received-power responses are presented in \figref{Fig:FF_to_FF_Sim_Meas}, where the continuously tunable RIS is compared with its binary counterpart for beam steering toward $\theta_{\mathrm{RX}}=\SI{30}{\degree}$ in the $\phi_{\mathrm{RX}}=\SI{90}{\degree}$ plane. A similar trend is observed in both the simulated and measured results. Although both implementations successfully generate the desired reflected beam, the binary RIS additionally exhibits a pronounced mirror-symmetric quantization lobe around $\theta_{\mathrm{RX}}=\SI{-30}{\degree}$, caused by the limited two-state phase quantization across the RIS aperture. In contrast, the continuously tunable RIS effectively suppresses this undesired lobe, concentrating the reflected energy in the desired direction.

The measured peak received power is further compared with an analytical
link-budget estimate based on the generalized Friis formulation for a
finite anomalous reflector~\cite{Kosulnikov_TAP_2023}, expressed as
\begin{subequations}\label{eq:linkbudget}
\begin{align}
P_\mathrm{r}
&=
P_\mathrm{t}G_\mathrm{TX}G_\mathrm{RX}
\left(
\frac{A}{4\pi d_1d_2}
\right)^2
\eta_\mathrm{eff}
\cos\theta_\mathrm{i}\cos\theta_\mathrm{r},
\label{eq:linkbudget_a}
\\
\intertext{or, equivalently, with all parameters in logarithmic form,}
P_\mathrm{r}
&=
P_\mathrm{t}
+
G_\mathrm{TX}
+
2G_\mathrm{RIS}
+
G_\mathrm{RX}
-
L_{\mathrm{FS},1}
-
L_{\mathrm{FS},2}
\nonumber\\
&\quad
+
10\log_{10}\!\left(
\eta_\mathrm{eff}
\cos\theta_\mathrm{i}\cos\theta_\mathrm{r}
\right),
\label{eq:linkbudget_b}
\end{align}
\end{subequations}
where $P_\mathrm{t}$ is the transmitted power, $G_\mathrm{TX}$ and $G_\mathrm{RX}$ are the transmitting and receiving horn gains, $A$ is the physical RIS aperture area, $d_1$ and $d_2$ are the Tx--RIS and RIS--Rx separations, and $\theta_\mathrm{i}$ and $\theta_\mathrm{r}$ are the incidence and reflection angles, respectively identified with $\theta_\mathrm{TX}$ and $\theta_\mathrm{RX}$. Moreover, $G_\mathrm{RIS}=10\log_{10}(4\pi A/\lambda^2)$ is the ideal RIS aperture gain, while $L_{\mathrm{FS},i}=20\log_{10}(4\pi d_i/\lambda)$ denotes the corresponding free-space path loss. The effective reflection efficiency is obtained from the measured array-level reflection loss reported in \secref{sec:NF_NF} as $\eta_\mathrm{eff}=|\Gamma_\mathrm{eff}|^2=10^{-5.41/10}$. 

For $P_\mathrm{t}=\SI{0}{dBm}$, $G_\mathrm{TX}=G_\mathrm{RX}=\SI{26.28}{dBi}$, $A=\SI{53.6}{mm}\times\SI{53.6}{mm}$,
$d_1=d_2=\SI{1.2}{m}$, $\theta_\mathrm{i}=\SI{0}{\degree}$, and $\theta_\mathrm{r}=\SI{30}{\degree}$, the ideal aperture gain and individual free-space path losses are $G_\mathrm{RIS}=\SI{24.98}{dBi}$ and $L_{\mathrm{FS},1}=L_{\mathrm{FS},2}=\SI{62.97}{dB}$, respectively. Including the measured reflection efficiency and the
$\SIrange{1.5}{2.0}{dB}$ finite-panel correction reported in~\cite{Kosulnikov_TAP_2023} yields a predicted received power of
$\SIrange{-31.5}{-31.0}{dBm}$, compared with the measured value of $\SI{-33.86}{dBm}$. The remaining difference of
$\SIrange{2.4}{2.9}{dB}$ is primarily attributed to the unmodeled insertion losses of the transmit and receive RF chains, together with calibration uncertainty, residual multipath, and the finite angular resolution of the manually positioned receiver. The agreement in both absolute level and angular response confirms that the continuous-phase beam-steering capability validated in the CATR is transferable to a practical far-field wireless link.

\subsection{Controlled Beam Splitting}\label{sec:Beam_Splitting}

Beyond single-beam steering, the proposed continuously programmable RIS enables programmable multi-beam wavefront synthesis through the superposition of independently synthesized single-beam phase distributions. Phase-only beam splitting has been reported for passive reflectarrays using a different synthesis approach~\cite{Young_AWPL_2026}, but its realization on a dynamic RIS remains uncommon. For two desired reflection directions $(\theta_{\mathrm{RX},1},\phi_{\mathrm{RX},1})$ and
$(\theta_{\mathrm{RX},2},\phi_{\mathrm{RX},2})$, let
$\Phi_{\mathrm{req},n}^{(1)}$ and $\Phi_{\mathrm{req},n}^{(2)}$ denote the corresponding phase distributions. A phase-only auxiliary synthesis function is then evaluated at each element as
\begin{subequations}\label{eq:BeamSplit}
\begin{equation}
S_n
=
\alpha_1 e^{j\Phi_{\mathrm{req},n}^{(1)}}
+
\alpha_2 e^{j\Phi_{\mathrm{req},n}^{(2)}},
\label{eq:BeamSplit_a}
\end{equation}
with the corresponding effective reflection phase
\begin{equation}
\Phi_{\mathrm{req},n}^{\mathrm{eff}}
=
\angle\left(S_n\right).
\label{eq:BeamSplit_b}
\end{equation}
\end{subequations}
Here, $\alpha_1$ and $\alpha_2$ are the synthesis weighting coefficients. Since
only their ratio affects \eqref{eq:BeamSplit_b}, $\alpha_1=1$ is selected and
$\alpha_2=\sqrt{P_2/P_1}$ specifies the nominal power ratio between the desired
beam powers $P_1$ and $P_2$. Only the argument of $S_n$ is retained; hence, the
surface remains phase-only and does not require element-level amplitude control.
Consequently, the realized beam-power ratio may differ from the nominal value.

\begin{figure}[t]
    \centering
    \setlength\figureheight{.20\textwidth}
    \setlength\figurewidth{.40\textwidth}
    \includegraphics{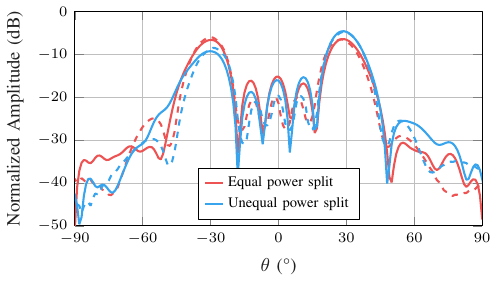}
    \caption{Experimental validation of multi-functional wavefront synthesis through programmable power redistribution among multiple beams. Dashed and solid lines denote the simulated and measured results, respectively.}
    \label{fig:Beam_splitting_equal_unequal}
    \end{figure}

The resulting phase distribution is implemented using the same unit-cell mapping procedure as for single-beam steering in \secref{sec:Analytical_Models} and is experimentally validated using the CATR setup of \figref{Fig:CATR_RIS_Meas_Setup}. As shown in \figref{fig:Beam_splitting_equal_unequal}, the RIS  simultaneously generates two beams at $\theta_{\mathrm{RX},1}=\SI{-30}{\degree}$ and
$\theta_{\mathrm{RX},2}=\SI{30}{\degree}$ in the
$\phi_{\mathrm{RX}}=\SI{0}{\degree}$ plane. For this symmetric configuration, all additional diffraction orders are evanescent. Both equal- and unequal-power cases are demonstrated. For the latter, the nominal ratio
$P_1:P_2=25:75$ corresponds to a power difference of approximately $\SI{4.77}{dB}$, while the measured difference of approximately $\SI{4.66}{dB}$ closely matches the nominal target despite the phase-only projection, demonstrating excellent agreement with the synthesized power distribution.
    
\begin{figure}[t]
    \centering
    \setlength\figureheight{.22\textwidth}
    \setlength\figurewidth{.40\textwidth}
    \includegraphics{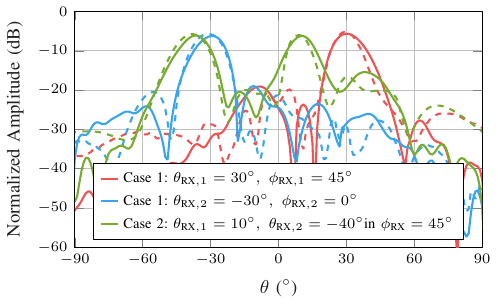}
    \caption{Programmable multi-beam generation in different angular directions and azimuthal planes using the proposed RIS. Measurement and simulation results are shown as solid and dashed lines, respectively.}
    \label{fig:Beam_splitting_multiple_cases}
    \end{figure}

The generality of the proposed approach is further demonstrated in \figref{fig:Beam_splitting_multiple_cases}: Case~1 generates two beams in different azimuthal planes, whereas Case~2 realizes asymmetric dual-beam synthesis. In Case~2, the phase-only superposition produces one additional
visible diffraction order near endfire ($\approx\SI{82}{\degree}$), where it is strongly attenuated by the
element-pattern roll-off. The good agreement between simulation and measurement in both cases confirms the capability of the proposed continuously programmable RIS to realize programmable multi-beam synthesis, including controllable power redistribution and simultaneous beam generation in different angular directions from a single continuously tunable aperture.

\subsection{Comparison and Discussion}\label{sec:Discussion}

\begin{table*}[!t]
\centering
\renewcommand\tabularxcolumn[1]{m{#1}}
\caption{Performance Comparison of the Proposed RIS With Previous Works}
\label{tab:RIS_Comparison}
\setlength{\tabcolsep}{2.0pt}
\renewcommand{\arraystretch}{1.12}
\scriptsize

\begin{tabularx}{\textwidth}{|>{\raggedright\arraybackslash}m{2.75cm}|*{7}{Z|}B|}
\hline

Reference
& \cite{Shekhawat_OJAP_2025}
& \cite{Shamim_TAP_2025}
& \cite{Kim_TAP_2023}
& \cite{Li_TAP_2024}
& \cite{Liang_TAP_2022}
& \cite{Wolff_JAP_2023}
& \cite{Ivashina_TAP_2025}
& \textbf{This Work}
\\
\hline\hline

Design frequency (\si{GHz})
& \num{27.2} & \num{25.9} & \num{28} & \num{37.5} & \num{3.15} & \num{31} & \num{28}
& \num{28}
\\
\hline

Phase control / quantization
& 1-bit & 2-bit & Continuous & Continuous & 3-bit & Continuous & Continuous
& Continuous; 3-/2-/1-bit emulation
\\
\hline

RF component type, Component no./ UC
& PIN, 1 & PIN, 2 & LC, 1 layer & LC, 1 layer & Varactor, 1 & Varactor, 1
& PIN, 1 \& Varactor, 1 & Varactor, 1
\\
\hline

UC periodicity ($\lambda_0$)
& $0.5$ & $0.39$ & $0.47$ & $0.5$ & $0.27\times0.29$ & $0.5$ & $0.42$ & \num{0.5}
\\
\hline

Phase coverage; states
& $180^\circ\pm20^\circ$; $2$-states
& $90^\circ\pm30^\circ$; $4$-states
& $230^\circ$/$260^\circ$; $6$-states used
& $338^\circ$ & $315^\circ$; $8$-states & $180^\circ$ & $360^\circ$ & \qty{300}{\degree}
\\
\hline

UC reflection loss (dB)
& $2$ & $0.8$ & $2$/$4$ & $10$ & $2$ & $7.33$ & $6$ sim.; $8.4$ meas.
& \num{4.6} sim.; \num{4.6}--\num{5.3} meas.
\\
\hline

Isolated-UC measurement, setup
& No & No & No & No & No & No & Yes; modified waveguide
& Yes; modified waveguide
\\
\hline

Array configuration
& $32\times32$ & $20\times20$ & $10\times10$ & $12\times12$ & $10\times10$ & $20\times20$
& No fabricated array & \num{96} elements
\\
\hline

NF--NF measurement
& Yes & Yes & Yes & Yes & No & Yes & N/A & Yes
\\
\hline

NF--FF beam-steering measurement
& No & Yes & No & Yes & No & No & N/A & Yes
\\
\hline

FF--FF measurement
& Yes & Yes & Yes & No & Yes & No & N/A & Yes
\\
\hline

Control integrated on RIS board
& No & No & No & No & No & No & N/A & Yes
\\
\hline

Power consumption$^{\ast}$:
tuning device(s)/UC (\si{\micro\watt});
array $+$ bias/drivers (\si{\watt});
total incl. controller (\si{\watt})
& $3200$;  $7.0$; $8^{\dagger}$
& $2100$; N/R; N/R         
& $0.06$; N/R; N/R
& $2.6$; N/R; N/R
& $0.5$; N/R; N/R
& $1.8$; N/R; N/R
& $2900$; N/A; N/A
& $\mathbf{1.8}$; $\mathbf{0.10^{\dagger}}$; $\mathbf{0.85^{\dagger}}$
\\
\hline

Full-aperture reconfiguration time
& N/R & N/R & N/R (slower) & $\sim$\SI{1}{\second} on / $\sim$\SI{6}{\second} off (meas.)
& N/R & N/R & N/A & $\bm{\sim}$\qty[detect-all]{50}{\milli\second} (est.)
\\
\hline

Additional array-level functionality
& 2-D beam steering & N/R
& Independent dual-polarization & 2-D beam steering & N/R & N/R
& N/A & Finite-bit emulation; controlled beam splitting
\\
\hline

\end{tabularx}

\vspace{1mm}
\begin{minipage}{0.995\textwidth}
\footnotesize
\textit{Notes:}
LC denotes liquid crystal. NF--NF, NF--FF, and FF--FF stand for near-field-to-near-field, near-field-to-far-field, and far-field-to-far-field, respectively. N/R and N/A denote not reported and not applicable, respectively.$^{\ast}$Values are estimated in this work from device datasheets; those marked $^{\dagger}$ are measured.
\end{minipage}
\end{table*}

\tabref{tab:RIS_Comparison} positions the proposed RIS against representative implementations based on the three dominant tuning mechanisms. PIN-diode surfaces provide fast and robust switching but remain limited to discrete 1-bit~\cite{Shekhawat_OJAP_2025} or 2-bit~\cite{Shamim_TAP_2025} phase states, making their radiation patterns susceptible to phase-quantization effects and the associated sidelobe and quantization-lobe levels. Liquid-crystal RISs enable continuous phase control~\cite{Kim_TAP_2023,Li_TAP_2024}, but may exhibit slow reconfiguration and comparatively high reflection loss~\cite{Li_TAP_2024}. Among varactor-based designs, the surface in~\cite{Liang_TAP_2022} operates in the sub-$\SI{6}{GHz}$ band with 3-bit control, while~\cite{Wolff_JAP_2023} demonstrates continuous mmWave steering over a limited phase range of $\SI{180}{\degree}$ and is characterized only under near-field conditions. The hybrid unit cell in~\cite{Ivashina_TAP_2025} provides full-$\SI{360}{\degree}$ phase coverage together with amplitude tunability, but employs two active elements per cell and is not demonstrated in a fabricated array. In comparison, the proposed RIS achieves $\SI{300}{\degree}$ of continuous phase tuning using a single varactor per unit cell. It is validated from waveguide-based unit-cell measurements through array-level NF--NF, NF--FF, and FF--FF experiments, while the same aperture also supports 3-/2-/1-bit emulation and controlled beam splitting.

A distinctive feature of the platform presented here is its low electrical-power requirement. Since the varactors are reverse-biased and draw negligible dc current, the complete $96$-element prototype, including the backside driver ICs, bias-generation network, and controller, consumes approximately $\SI{0.85}{\watt}$. Under the respective reported system boundaries, this is lower than the $\SI{8}{\watt}$ including control electronics reported in~\cite{Shekhawat_OJAP_2025} and the approximately $\SI{2}{\watt}$ required for PIN biasing alone in~\cite{Shamim_TAP_2025}. The associated reconfiguration speed is governed by the settling time of the PWM-to-dc filters rather than by the varactors themselves. Under the equivalent first-order low-pass model, each tuning channel is described by an effective resistance of $R_\mathrm{eff}=\SI{150}{\kilo\ohm}\parallel\SI{330}{\kilo\ohm}\approx\SI{103}{\kilo\ohm}$ and $C=\SI{100}{\nano\farad}$, yielding a nominal time constant of $\tau=R_\mathrm{eff}C\approx\SI{10.3}{\milli\second}$. For a first-order response, the settling time for a residual bias error of $\SI{1}{\percent}$ is
$t_\mathrm{s}=-\tau\ln(0.01)\approx\SI{47.5}{\milli\second}$.
Including the comparatively small PWM-update and serial-transfer latencies, the estimated full-aperture reconfiguration time is approximately $\SI{50}{\milli\second}$, corresponding to an update rate of about $\SI{20}{\hertz}$. All $96$ channels are latched simultaneously, and their bias networks settle in parallel; hence, the dominant analog settling time does not increase with the number of simultaneously updated elements, although the serial-transfer latency increases with the number of cascaded drivers. Faster operation could be obtained by reducing the filter time constant or adopting dedicated digital-to-analog converters (DACs); however, the latter can substantially increase system power, as illustrated by the $\SI{12.8}{\watt}$ consumption reported for a comparable $144$-element varactor reflectarray~\cite{Fischer_EuCAP_2025}. The present implementation therefore provides a deliberate speed--power trade-off while remaining one to two orders of magnitude faster than the measured liquid-crystal response in~\cite{Li_TAP_2024}.

The principal trade-offs of the single-varactor implementation are the absence of independent amplitude and phase control, moderate reflection loss, and limited fractional bandwidth. These limitations are balanced by continuous phase tuning enabled by the calibrated PWM-to-bias-voltage mapping, avoidance of quantization lobes associated with native low-bit operation, and full integration of the bias electronics on the RIS board. The same hardware can also emulate 3-/2-/1-bit phase control and realize controlled beam splitting without modifying the aperture. Finally, the architecture is explicitly scalable, with dedicated edge interconnects allowing multiple RIS modules to be daisy-chained and additional driver ICs enabling the bias channels to be extended in increments of $24$, thereby supporting larger apertures without changing the basic control topology.

\section{Conclusion}\label{sec:Conclusion}
This paper established and experimentally validated a device-to-system modeling framework that links measured varactor-based unit-cell characteristics to the beamforming performance of a continuously tunable mmWave RIS operating at a design frequency of $\SI{28}{GHz}$. Building upon an experimentally validated unit-cell model, a practical $96$-element RIS with integrated analog bias control and a unified analytical framework incorporating feed-horn illumination, finite phase availability, and reflection losses were developed to accurately predict the measured beam-steering performance. Experimentally, the proposed RIS platform achieved continuous beam steering for all investigated angles within $|\theta_{\mathrm{RX}}| \leq \SI{45}{\degree}$ across three azimuthal planes, with a maximum deviation of approximately $\SI{2}{\degree}$. Furthermore, an estimated full-aperture reconfiguration time of approximately $\SI{50}{ms}$ can be achieved at a total power consumption of only $\SI{0.85}{W}$. Additionally, a multi-beam synthesis leveraging the continuous phase control has been demonstrated. The resulting agreement among analytical modeling, full-wave simulations, and experimental characterization demonstrates that practical hardware nonidealities can be consistently captured from the device level to the system level.

Beyond beam steering, the proposed platform enables a unified experimental investigation of continuously tunable RIS operation, including aperture characterization, finite-bit phase quantization, practical wireless-link assessment, and programmable multi-beam wavefront synthesis using a common hardware architecture. Rather than representing isolated demonstrations, these investigations collectively establish a comprehensive experimental methodology for evaluating continuously tunable RISs under realistic operating conditions.

The presented methodology provides a rigorous foundation for the development and experimental validation of future continuously tunable mmWave RISs, bridging measured device behavior with analytical modeling and programmable electromagnetic wavefront synthesis.


\bibliographystyle{IEEEtran}
\bibliography{IEEEabrv,Reference}
\end{document}